%% file: top2016.tex
\documentclass[12pt]{article}
\usepackage{graphicx}

\newcommand\pubnumber{\ }
\newcommand\pubdate{\today}

\def\institute{CERN, EP department, CH-1211 Geneva 23, Switzerland}
\def\support{\footnote{On behalf of the ATLAS Collaboration}}

\def\Title#1{\begin{center} {\Large #1 } \end{center}}
\def\Author#1{\begin{center}{ \sc #1} \end{center}}
\def\Address#1{\begin{center}{ \it #1} \end{center}}

\newcommand\pubblock{\rightline{\begin{tabular}{l} \pubnumber\\
         \pubdate  \end{tabular}}}
\newenvironment{Abstract}{\begin{quotation}  }{\end{quotation}}
\newenvironment{Presented}{\begin{quotation} \begin{center} 
             PRESENTED AT\end{center}\bigskip 
      \begin{center}\begin{large}}{\end{large}\end{center} \end{quotation}}

\input econfmacros.tex

\newcommand{\ttbar}{\mbox{$t\bar{t}$}}
\newcommand{\qqbar}{\mbox{$q\bar{q}$}}
\newcommand{\xtt}{\mbox{$\sigma_{\ttbar}$}}

\newcommand{\sxwt}{$\sqrt{s}=7$\,TeV}
\newcommand{\sxyt}{$\sqrt{s}=13$\,TeV}

\newcommand{\pt}{\mbox{$p_{\rm T}$}}
\newcommand{\etmiss}{\mbox{$E_{\rm T}^{\rm miss}$}}

\newcommand{\aconfref}[3]{ATLAS Collaboration, ATLAS-CONF-#2}
\newcommand{\apubref}[3]{ATLAS Collaboration, ATLAS-PHYS-PUB-#2}

\newcommand{\arefpaper}[3]{ATLAS Collaboration, #2, arXiv:#3}

\newcommand{\aplb}[3]{\arefpaper{#1}{Phys.\ Lett.\ B#2}{#3}}

\newcommand{\aepjc}[3]{\arefpaper{#1}{Eur.\ Phys.\ J.\ C#2}{#3}}

\begin{document}
\begin{titlepage}
\pubblock

\vfill
\Title{Physics objects for top quark physics in ATLAS}
\vfill
\Author{ Richard Hawkings\support}
\Address{\institute}
\vfill
\begin{Abstract}
Top quark physics measurements performed using data from the ATLAS
detector at the LHC rely on efficient reconstruction and precise calibration
of leptons, jets and missing transverse energy. A review of the techniques
used to reconstruct such objects is given, with an emphasis on the uncertainties
achieved for energy calibration and efficiency measurements, illustrated 
with their impact on key top quark physics results.
\end{Abstract}
\vfill
\begin{Presented}
$9^{th}$ International Workshop on Top Quark Physics\\
Olomouc, Czech Republic,  September 19--23, 2016
\end{Presented}
\vfill
\end{titlepage}
\def\thefootnote{\fnsymbol{footnote}}
\setcounter{footnote}{0}

\section{Introduction}

The study of events containing top quark pairs (\ttbar) 
or single top quarks forms a key part of the ATLAS \cite{atlasdet}
proton--proton ($pp$) physics program at
the CERN Large Hadron Collider (LHC).
Within the Standard Model, 99.8\,\% of top quarks decay into a $W$ boson
and $b$ quark ($t\rightarrow Wb$), so the final states are determined
by the decay modes of the $W$ boson: leptonic decays $W\rightarrow\ell\nu$
with the lepton $\ell$ being an electron, muon or tau, or hadronic decays
$W\rightarrow\qqbar$ with the quarks giving rise to two collimated jets
of hadrons in the detector. The reconstruction of leptons, jets (including
those tagged as originating from $b$ quarks), and the missing transverse
energy from neutrinos produced in $W\rightarrow\ell\nu$ decays, is therefore
crucial in fully exploiting the potential of the LHC for top quark physics.

Table~\ref{t:samples} shows the data samples delivered by the LHC to ATLAS
so far at different centre-of-mass energies $\sqrt{s}$; nearly 6M \ttbar\
pairs were produced during Run-1 ($\sqrt{s}=$7--8\,TeV), and over 30M are now
available at \sxyt\ from Run-2. These samples come with increasing numbers 
of simultaneous $pp$ collisions per bunch-crossing (pileup), 
posing significant challenges for the
reconstruction of the high transverse-momentum (\pt) objects produced 
in top quark decays.

\begin{table}[h]
\begin{center}
\begin{tabular}{lc|cc|cc}  
Year & $\sqrt{s}$ (TeV) & $<\mu>$ & $L_{\rm int}$ (fb$^{-1}$) & \xtt\ (pb) &
$N(\ttbar)$ \\ \hline
2011 & 7 & 9 & 4.6 & 170 & 800k \\
2012 & 8 & 20 & 20.2 & 250 & 5M \\
2015 & 13 & 14 & 3.2 & 830 & 2.6M \\
2016 & 13 & 25 & 33.3 & 830 & 28M \\
\end{tabular}
\caption{Data samples available for top physics studies in ATLAS,
showing the year and centre-of-mass energy $\sqrt{s}$, average
number of interactions per crossing $<\mu>$, integrated luminosity $L_{\rm int}$,
\ttbar\ production 
cross-section \xtt, and approximate number of \ttbar\ events in each sample.
\label{t:samples}
\vspace{-14mm}
}
\end{center}
\end{table}

\section{Leptons}

Electrons are identified from a shower in the electromagnetic calorimeter, 
spatially matched to a track reconstructed in the inner detector 
\cite{elec7}. The major backgrounds from misidentified
hadrons and photon conversions are reduced via cuts on calorimeter shower
shapes, matching between the calorimeter energy and track momentum, the
detection of transition radiation in the TRT straw-tube tracker, and the
presence of a track hit in the first layer of the pixel detector, giving 
efficiencies of 80-95\,\% for electrons with $\pt>25$\,GeV.
Muons are reconstructed from tracks found independently in the inner detector
and muon spectrometer, which are then tested for compatibility and combined
with a global track fit. This gives an efficiency above 98\,\% for $\pt>25$\,GeV
whilst strongly suppressing the background from
$\pi/K$ decays in flight and punch-through of showers from the hadronic
calorimeter \cite{muon13}.

\begin{figure}[tp]
\hspace{-10mm}
\includegraphics[width=80mm]{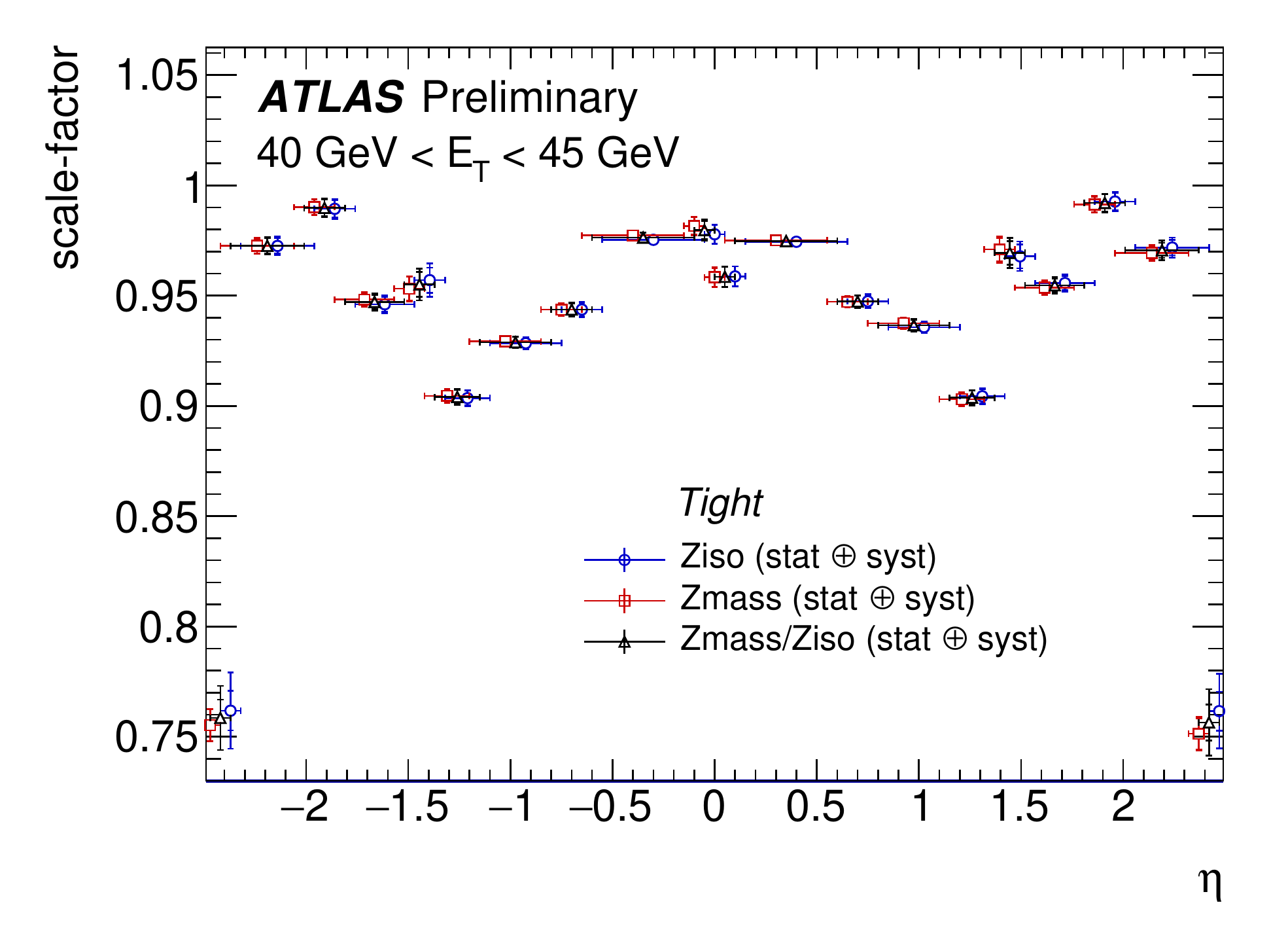}
\includegraphics[width=85mm]{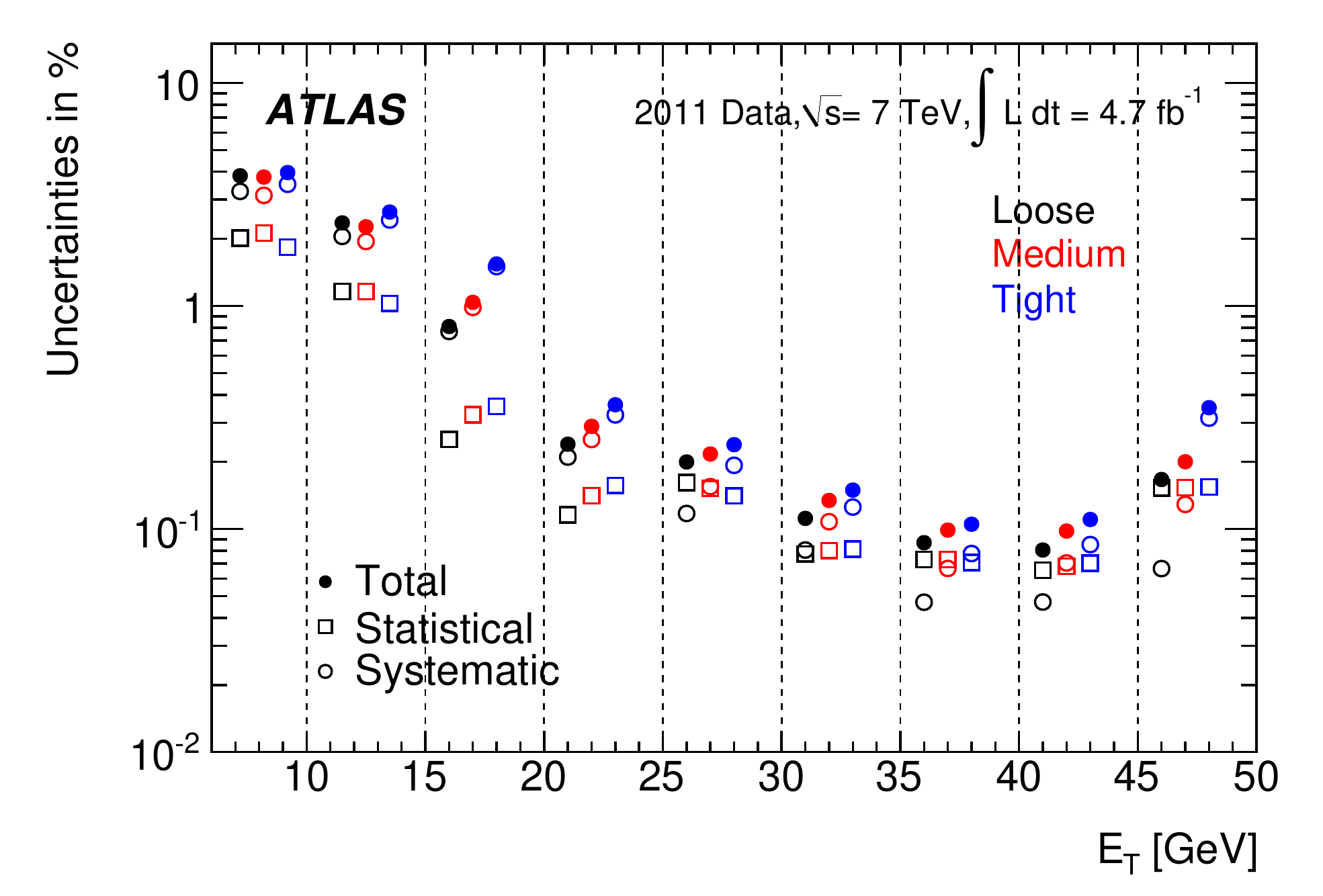}
\vspace{-10mm}
\caption{\label{f:elec}(left) Electron identification efficiency scale factors
measured from $Z\rightarrow ee$ tag-and-probe analysis at \sxyt\ \cite{elec13};
(right) uncertainties on electron efficiency scale factors at \sxwt\ 
\cite{elec7}.}
\end{figure}

\begin{figure}[tp]
\vspace{-10mm}
\includegraphics[width=64mm]{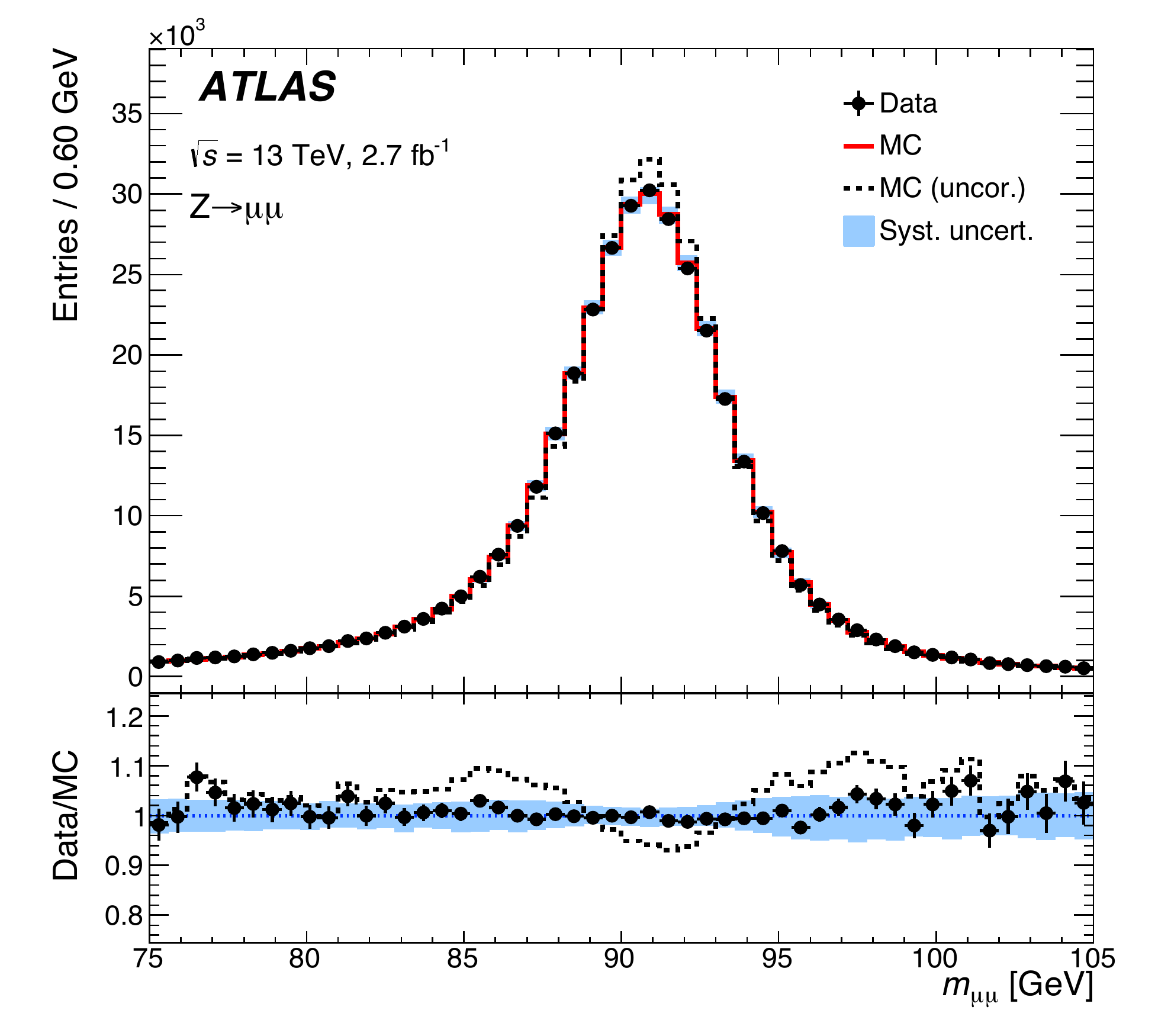}
\includegraphics[width=80mm]{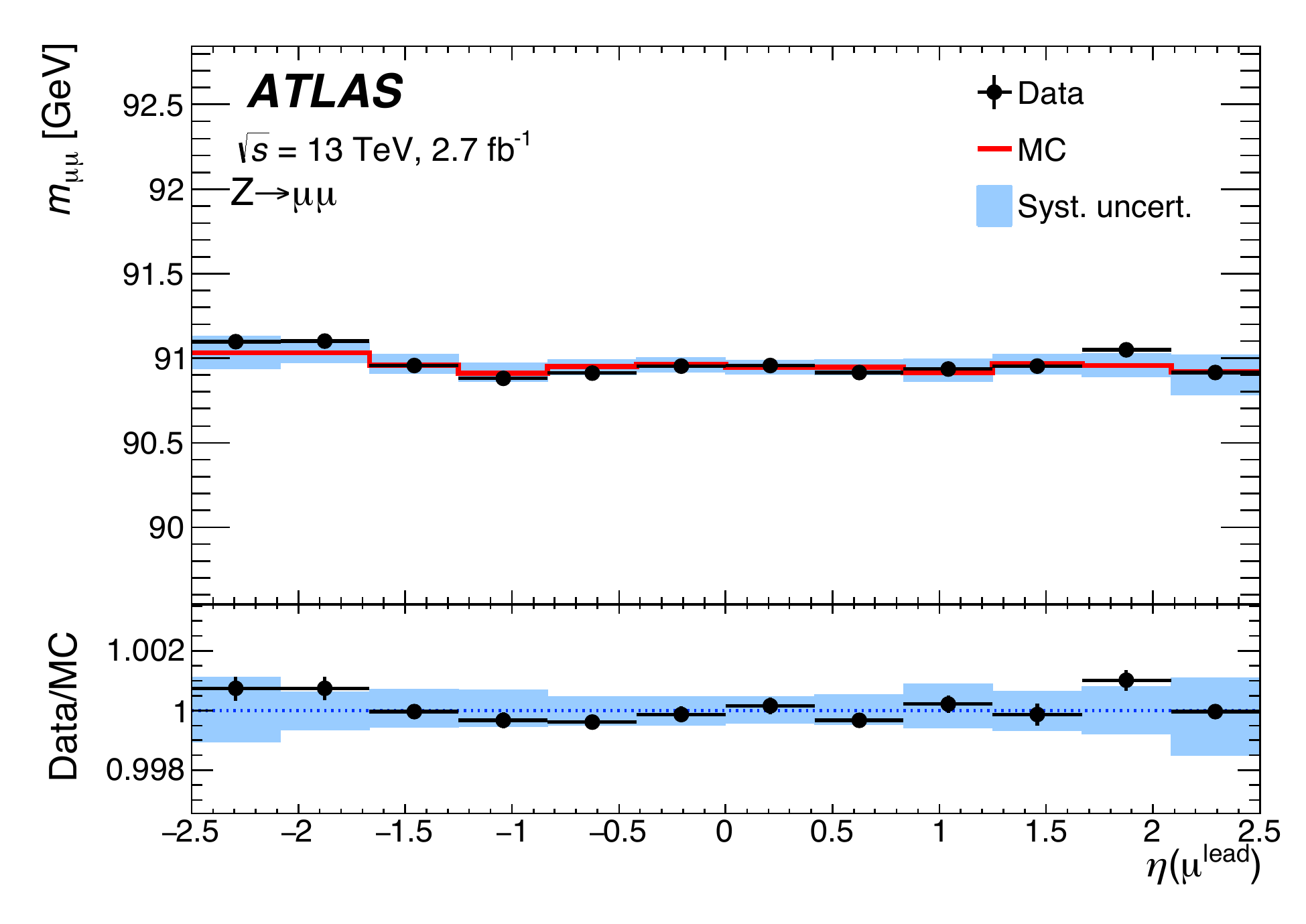}
\vspace{-4mm}
\caption{\label{f:muon}(left) Di-muon invariant mass distribution for selected
$Z\rightarrow\mu\mu$ events compared to uncorrected and corrected simulation 
with systematic uncertainty; (right) stability of mean reconstructed mass
vs. $\eta$ of the highest \pt\ muon \cite{muon13}.}
\end{figure}

The identification efficiencies for both lepton types are measured using 
the tag-and-probe technique applied to $Z\rightarrow\ell\ell$, 
$W\rightarrow e\nu$ and
$J/\psi\rightarrow\ell\ell$ ($\ell=e$, $\mu$) decays, requiring one lepton (the tag)
to pass tight trigger and identification requirements, and using the resonance
mass distribution to determine the background in the other (probe) lepton 
sample without applying the identification requirements. The 
efficiencies are measured as functions of lepton \pt\ and pseudo-rapidity
$\eta$, and  expressed as scale factors with respect to the predictions from 
simulation. For electrons, the scale factors are typically within 5\,\% of
unity (except in regions with large amounts of material in front of
the electromagnetic calorimeter) and measured
to a precision well below 1\,\% (Figure~\ref{f:elec}). For muons the scale
factors are within 1\,\% of unity and measured with similar precision.
The same $Z\rightarrow\ell\ell$ samples are used to provide in-situ corrections
to the electron energy and muon momentum scales, using the known value of 
the $Z$-boson mass. The width of the reconstructed $Z\rightarrow\ell\ell$
mass distribution is sensitive to the energy/momentum resolution, and used
to adjust the resolution in simulation to better model the data, as shown
for muons in Figure~\ref{f:muon}. The uncertainties on the energy/momentum 
scales are typically smaller than $10^{-3}$ in the region close to the $Z$ 
resonance, but some extrapolation to higher energy/momentum is needed
to cover the leptons produced in top quark decays.

Leptons from $W$ decays in top events are typically isolated from other
hadronic activity, so the sums of calorimeter energy deposits 
(after correction for pileup) and track momenta close to the leptons are 
required to be small. These selections reject leptons from semi-leptonic 
decays of heavy flavour hadrons.
Leptons also provide efficient 
triggers for selecting top events online, with efficiencies for 
$\pt>25$\,GeV above 90\,\% for electrons and 70-85\,\% for muons, limited by
the geometrical coverage of the trigger chambers. Dilepton events are typically
selected with a logical OR of single lepton triggers, providing a robust
trigger with 99\,\% per-event efficiency, and small systematic uncertainties,
again measured using $Z$ tag-and-probe techniques. 

The uncertainties from limited knowledge of lepton efficiencies and calibration
in the ATLAS \ttbar\ inclusive cross-section measurements from
$e\mu$ dilepton events at $\sqrt{s}=7$, 8 and 13\,TeV \cite{attx}
are shown in Table~\ref{t:xsecsyst}. These are all well below 1\,\% (even 
in the less mature \sxyt\ analysis) and significantly smaller than the
leading uncertainties from \ttbar\ modelling and luminosity measurement.

\section{Jet, $b$-tagging and missing transverse energy}

\begin{figure}[tp]
\vspace{-5mm}
\includegraphics[width=82mm]{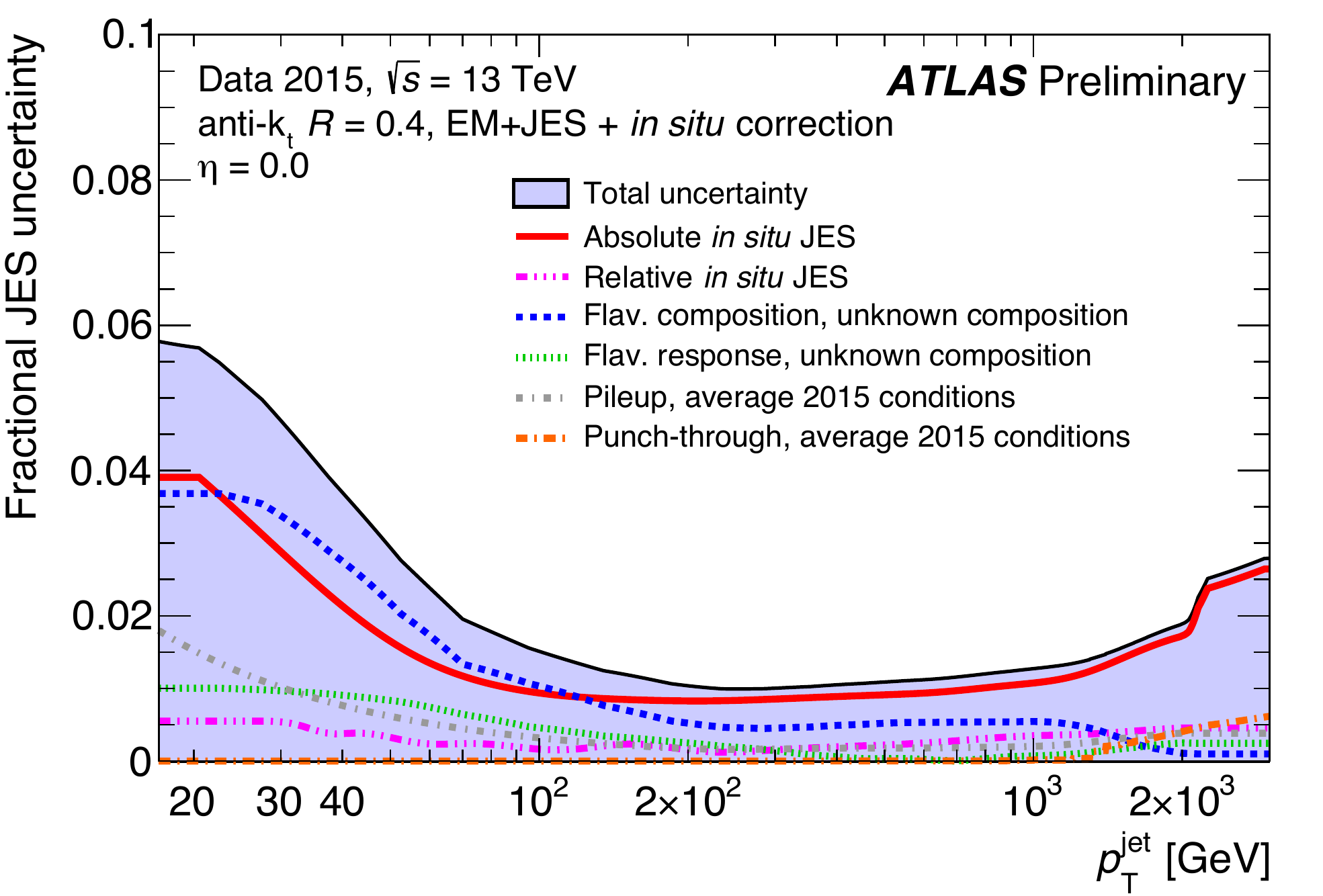}
\hspace{5mm}
\includegraphics[width=60mm]{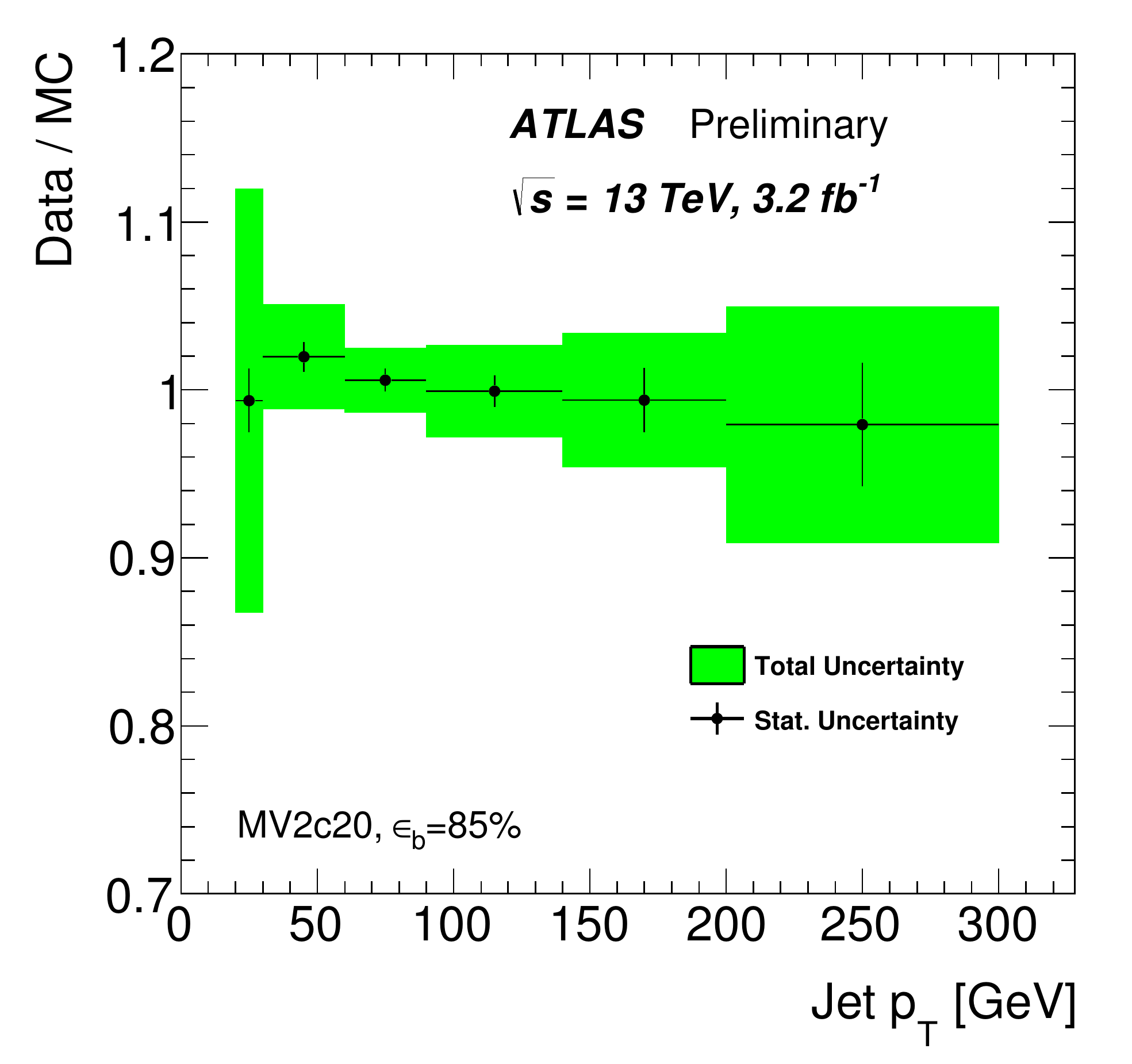}
\vspace{-4mm}
\caption{\label{f:jetb}(left) Fractional jet energy scale uncertainty as
a function of jet \pt\ for jets at $\eta=0$ in 2015 data at \sxyt,
illustrating the contributions of the various components \cite{jetpub};
(right) $b$-tagging efficiency scale factors with statistical and systematic
uncertainties measured using \ttbar\ events in \sxyt\ data \cite{btagsf}.}
\end{figure}

The outgoing quarks and gluons from the hard-scattering collision are 
reconstructed as collimated jets of hadrons in the detector. ATLAS
uses the anti-$k_T$ jet algorithm applied to topological clusters of energy
deposits in the calorimeters. The simulation-based jet energy scale calibration
is augmented with data-based corrections exploiting energy balance in
photon+jet, $Z$+jet and multijet events \cite{jetpub}. The resulting systematic
uncertainties as a function of jet \pt\ are shown in Figure~\ref{f:jetb} (left);
uncertainties of below 2\,\% for $\pt=100$\,GeV have already been achieved
in Run-2 data, with up to a factor two better being achieved in Run-1. 
Jet energy
scale uncertainties are typically amongst the leading ones for measurements
of jet activity in top events, and in measurements of the top mass,
contributing e.g. 0.6\,GeV to the top mass uncertainty in the lepton+jets
measurement at \sxwt\ \cite{topmass7}.

\begin{table}[tp]
\begin{center}
\begin{tabular}{l|ccc}
Uncertainty source (\%) & 7\,TeV & 8\,TeV & 13\,TeV \\ \hline
Electron efficiency & 0.13 & 0.41 & 0.3 \\
Electron scale/resolution & 0.22 & 0.51 & 0.2 \\
Electron isolation & 0.59 & 0.30 & 0.4 \\
Muon efficiency & 0.30 & 0.42 & 0.4 \\
Muon scale/resolution & 0.14 & 0.02 & $<0.05$ \\
Muon isolation & 0.44 & 0.22 & 0.3 \\ 
Lepton trigger & 0.19 & 0.16 & 0.2 \\
\end{tabular}
\caption{\label{t:xsecsyst}Lepton-related relative uncertainties (in \%) 
on the measurements of the \ttbar\ production cross-section at 
$\sqrt{s}=7$,8 and 13\,TeV \cite{attx}.}
\end{center}
\end{table}

At high invariant masses of the \ttbar\ system, the top quarks become
highly boosted, and the jets from hadronic top decays
$t\rightarrow bq\bar{q}$ can no longer be resolved. Jet substructure techniques,
in which a large radius ($R=1.0$) jet capturing all the top quark 
decay products is 
broken down to reveal the mass of the heavy object within, whilst removing the
soft contributions from QCD radiation and pileup, are  essential
in measuring differential cross-sections at high top quark \pt\ \cite{boosttop}.

The identification of jets likely to have originated from $b$-quarks
plays an important role in isolating events containing top quarks (due to
the dominant $t\rightarrow Wb$ decay), and assigning jets to top quark decay
products when performing kinematic fits. With the addition of the Inner B-Layer
pixel detector at $r=33$\,mm from the collision point, 
the $b$-tagging efficiency has been boosted in Run-2 compared to Run-1 
by around 10\,\% for a similar rejection of light quark, gluon
and charm jet background \cite{btagpub}. The $b$-tagging efficiency is 
also measured in data by exploiting \ttbar\ events, allowing efficiency
scale factors to be derived with a precision of 2--3\,\% for jets
in the 50--150\,GeV \pt\ range \cite{btagsf},
 as shown in Figure~\ref{f:jetb} (right).

The neutrinos produced in the leptonic decays of $W$ bosons from top quark
decays escape detection and give rise to missing transverse energy (\etmiss), 
i.e. imbalance in the vector sum of transverse momentum of all objects 
in the final state. The \etmiss\ resolution is affected by the
resolution of the soft term, i.e. the residual 
energy not clustered into jets or electrons. 
In Run-2, the contribution of pileup
to this soft term is minimised by replacing calorimeter energy clusters
by tracks reconstructed in the inner detector and associated to the
primary vertex from the hard-scattering collision rather than those
from pileup interactions. The reconstructed \etmiss\ can be used
to separate top quark events from multijet background, and to help 
reconstruct the kinematics of the top quark(s) in the event.

\section{Outlook}

Top physics studies rely on measurements of many of the physics objects
ATLAS can reconstruct. The detector, reconstruction and calibration
procedures are working well at \sxyt, and uncertainties are beginning to
approach those achieved with the Run-1 data. Top analyses are typically
limited by the systematic uncertainties related to jets, rather than leptons
which can be precisely calibrated using $Z\rightarrow\ell\ell$ decays. 
Jet substructure techniques are also beginning to bear fruit, and ATLAS
looks forward to more data and top quark physics results in the next few
years.

\end{document}

%% file: econfmacros.tex
\def\beq{\begin{equation}}
\def\eeq#1{\label{#1}\end{equation}}
\def\eeqn{\end{equation}}

\def\beqa{\begin{eqnarray}}
\def\eeqa#1{\label{#1}\end{eqnarray}}
\def\eeqan{\end{eqnarray}}

\let\bar=\overbar

\def\Dslash{\not{\hbox{\kern-4pt $D$}}}
\def\dslash{\not{\hbox{\kern-2pt $\del$}}}

\def\msb{{\bar{\ssstyle M \kern -1pt S}}}

